\documentclass[12pt]{article}
\usepackage{graphicx}
\usepackage{amssymb}
\begin{document}
\begin{center}
{\Large\bf Interacting holographic dark energy models: A general approach }\\[15mm]
S. Som
\footnote{Meghnad Saha Institute of Technology, Nazirabad, East
Kolkata Township,\\Kolkata 700 107,India;\indent email:
sumitsom79@yahoo.com},
A. Sil \footnote{St.Paul's C. M. College, 33/1 Raja Rammohan Roy Sarani, Kolkata 700 009, India;\\
\indent email: amitavadrsil@rediffmail.com}\\[5mm]

{\em Relativity and Cosmology Research Centre,\\Department of Physics, Jadavpur University,\\
Kolkata - 700 032, India.} \\[20mm]
\end{center}

\pagestyle{myheadings}
\newcommand{\be}{\begin{equation}}
\newcommand{\ee}{\end{equation}}
\newcommand{\bea}{\begin{eqnarray}}
\newcommand{\eea}{\end{eqnarray}}

\begin{abstract}
Dark energy models inspired by the cosmological holographic principle are studied in homogeneous isotropic spacetime with a general choice for the dark energy density $\rho_d=3(\alpha H^2+\beta\dot{H})$. Special choices of the parameters enable us to obtain three different holographic models, including the holographic Ricci dark energy(RDE) model. Effect of interaction between dark matter and dark energy on the dynamics of those models are investigated for different popular forms of interaction. It is found that crossing of phantom divide can be avoided in RDE models for $\beta>0.5$ irrespective of the presence of interaction. A choice of $\alpha=1$ and $\beta=2/3$ leads to a varying $\Lambda$-like model introducing an IR cutoff length $\Lambda^{-1/2}$. It is concluded that among the popular choices an interaction of the form $Q\propto H\rho_m$ suits the best in avoiding the coincidence problem in this model.

\end{abstract}

\vspace{0.5cm} Key Words: Holographic principle, Phantom barrier, Holographic Ricci \\
\indent dark energy, Varying $\Lambda$
\vspace{0.5cm}

\section{Introduction}
Observational evidence behind the claim that the Universe is presently going through an accelerated phase of expansion\cite{accln} is now so strong that the main focus of present day observational cosmology has shifted towards answering questions such as when has the acceleration begun or what might be the true nature of the agent driving such acceleration. The confirmation of this acceleration was provided recently by the third year data of the WMAP mission\cite{spergel}. Also the data from high redshift supernovae type Ia\cite{riess}, the cosmic microwave background radiation\cite{hanany}, the large scale structure\cite{colless}, the integrated Sachs-Wolfe effect\cite{boughn}, and weak lensing\cite{contaldi} endorse it undoubtedly. On the basis of observational data the standard cosmology suggests that more than 70\% of our Universe is made up of a form of energy that exerts a negative pressure and drives the cosmic acceleration. The unknown nature of such entity brings it a name `Dark Energy'. On the other hand, a little less than 30\% of the content of the Universe is gravitating matter, but most of it is non-baryonic and is called `Dark Matter' again due to its unknown nature. Thus our present understanding of the Universe is mostly phenomenological. Most of the theoretical investigations assume the cosmic fluid made up of dark matter and dark energy evolving independently. In a Robertson-Walker space-time this implies that pressureless dark matter energy density varies as $a^{-3}$ where $a$ is the scale factor of the metric. The evolution of dark energy density then solely depends on its equation of state, search of which is a major concern in present day observational cosmology. Any possible interaction between dark matter and dark energy however can not be ruled out. There exist no strong theoretical arguments nor enough observational evidences negating the possibility of such an interaction. In fact by ignoring the possibility we may misinterpret data concerning dark energy equation of state. On the other hand general approach to include interaction has the benefit that they admit non-interacting limit. In other words inclusion of interaction leads to corrections to the non-interacting configuration. Dark matter-dark energy interacting models were introduced by Wetterich\cite{wett} and are now common in literature\cite{zimdahl,farr}. In fact an appropriate choice of the interaction rate may help in resolving the question as to why are the densities of dark matter and dark energy precisely of the same order in the present epoch: a problem commonly referred to as the `coincidence problem'.

In absence of any concrete knowledge about the nature of dark energy, the list of proposed candidates for the same is already long and getting longer day by day. It is usually parameterized by an equation of state of the form $w_{d} = p_d/\rho_d$. From Friedmann equation one can see that a value of $w_{\it eff} < -1/3$ is required for accelerated cosmic expansion, where $w_{\it eff}$ is the effective equation of state for the cosmic fluid. The most natural and simplest choice for dark energy candidate could have been the Cosmological Constant $\Lambda$ with an equation of state $w _{d}=-1$. But if one believes that vacuum energy is the origin of it then one fails to find any mechanism to obtain a value of $\Lambda$ that is 120 orders of magnitude less than the theoretical prediction to be consistent with observation. Attempts were made by introducing the concept of varying $\Lambda$ to explain observation but with same equation of state $w _{d}=-1$\cite{varying lambda}. Possibilities of evolving $w _{d}$ were also explored in many dynamical dark energy models. Primary candidates in this category are scalar field models such as Quintessence\cite{quintessence} and K-essence\cite{k essence}. A common example to quintessence is the energy of a slowly evolving scalar field that has not yet reached the minimum of its potential $V (\phi)$, similar to the inflaton field used to explain the inflationary Universe. In quintessence models\cite{turner} the dark energy density decreases with the scale factor $a(t)$ as $\rho_{d}\propto a^{-3(1+w_{d})}$ where $-1 < w_d < -1/3$. An exotic form of dark energy with $w_{d} < -1$, named the phantom energy, has also been proposed\cite{Caldwell}. The phantom energy violates the energy conditions and has an infinitely increasing energy density. However, recent fits to observational data mildly favor an evolving equation of state crossing the phantom divide and several such models have been proposed in the literature\cite{phantom}. There exist other classes of `modified matter' dynamical dark energy models, such as Chaplygin gas\cite{chaplygin} and also many `modified general relativity' theories such as $f(R)$ gravity\cite{f R}, Scalar-Tensor theory\cite{st} inspired models and Braneworld models\cite{brane} in the list. A few comprehensive and recent review articles on dark energy may be of interest in this regard\cite{darkenergy}.
\par A new alternative to the solution of dark energy problem may be found in the `Holographic Principle'. The whole idea that `three dimensional world might be considered as an image of data that can be stored on a two dimensional projection much like a holographic image', began to develop when Bekenstein\cite{bekenstein} noticed the similarity in behavior of the black hole(BH) area and its entropy. The striking similarity that both tend to increase irreversibly led him to adopt a thermodynamical approach to BH physics. He showed from dimensional argument that BH entropy is equal to its area in Planck unit times a dimensionless constant of order unity. To reconcile this proposal with quantum mechanics t'Hooft\cite{hooft} introduced the holographic principle which was later extended by Susskind\cite{susskind} to string theory. According to this principle the entropy of a system scales not with its volume, but with its surface area\cite{susskind,bousso}. In other words, the degrees of freedom of a spatial region reside not in the bulk but only at the boundary of the region and the number of degrees of freedom per Planck area is no greater than unity. A cosmological version of this principle was later proposed by Fischler and Susskind\cite{fischler} which states: at any time during cosmological evolution, the gravitational entropy within a closed surface should be always larger than the particle entropy that passes through the past light-cone of that surface. Thus the holographic principle will set an upper bound on the entropy of the universe. In their work Cohen et al.\cite{cohen} pointed out that the proposal of Bekenstein that entropy of black-hole scales with its area rather than its volume, essentially suggests that quantum field theory breaks down in large volume.  To reconcile this breakdown with the success of local quantum field theory in describing observed particle phenomenology, they proposed that a short distance (UV) cutoff must be related to a long distance (IR) cutoff so that an effective field theory could be a good description of nature. If the quantum zero-point energy density $\rho_{vac}$ is caused by a short distance cut-off, the total energy in a region of size L should not exceed the mass of a black hole of the same size, i.e $L^3\rho_{vac}\leq LM_p^2$. Thus, if we take the whole Universe into account, then the vacuum energy related to this holographic principle may be viewed as dark energy, usually called holographic dark energy. The largest L allowed is the one saturating this inequality, therefore
\[
\rho_d=3c^2M_p^2L^{-2}
\]
where $c^2$ is a numerical constant, $M_p^2=8\pi G$ and $\rho_d$ represents the dark energy density.

\par In the context of the dark energy problem, the cutoff length was initially proposed\cite{cohen,hubra} to be $L=H^{-1}$ which implies that the unknown vacuum energy density $\rho_d$ is proportional to the square of the Hubble radius $\rho_d\propto H^2$. They argue that this proposed IR bound has no conflict with any current experimental success of quantum field theory, but explains why conventional effective field theory estimates of the cosmological constant fail so miserably. This in principle solves the fine tunneling problem, but the equation of state for dark energy is found to be zero and hence does not contribute to the present accelerated expansion. Use of particle horizon\cite{bousso,fischler} as the length scale, as S. Hsu\cite{hsu} and M. Li\cite{li} pointed out, can neither explain the present acceleration as it gives an equation of state parameter higher than $-1/3$. The choice of future event horizon as the length scale is able to produce the desired accelerated regime\cite{li}. However, R. Cai\cite{cai} pointed out that this proposal suffer from drawbacks concerning causality violation. He argued that if the present value of dark energy is to be determined by the future evolution of the Universe then it poses a serious challenge to the concept of causality. It is also very unlikely that a local quantity like the density of dark energy will get determined by the event horizon, which is a global concept of space-time.  Moreover, in the case of a spatially flat Friedmann-Robertson-Walker Universe, one can show that the future event horizon will exist if and only if the Universe is accelerating. So it is indeed perplexing that in order to interpret the cosmic acceleration, the holographic dark energy model with future event horizon as cut-off has itself presumed the acceleration. Subsequently, other holographic models of dark energy based on the different choices of IR cutoff were also proposed. For example the choice of the Ricci scalar as the scale of the Universe gives rise to the holographic Ricci dark energy model(RDE)\cite{gao}. Suwa-Nihei\cite{suwa} and Fu et al.\cite{fu} extended the RDE model to include the possible interactions between dark energy and cold dark matter. In these models, the spatially flat Universe is found to enter in to an accelerated phase of expansion in the recent past but the dark energy equation of state crosses the phantom barrier. Crossing the phantom divide is however an undesirable feature in holographic dark energy models. This is because in deriving the holographic bound it is necessary that the positive energy condition be satisfied. Phantom like dark energy which violates the positive energy condition is therefore incompatible with the holographic idea. Hence it is desired that the equation of state for dark energies of holographic origin must satisfy the condition $w_d\geq-1$\cite{zimdahl,bak, seta1}. A detailed analysis of various holographic dark energy models is available in \cite{delcampo}. Holographic dark energy in non-flat Universe has been extensively studied by Setare\cite{seta1, seta2}.

\par From purely dimensional grounds Granda and Oliveros\cite{granda} proposed a new infrared cut-off for the holographic density which includes time derivative of the Hubble parameter. They suggested a holographic dark energy density of the form $\rho_d\sim\alpha H^2+\beta \dot{H}$ described to be the simplest case of a general $f(H, \dot{H})$ holographic density in the FRW background. In favor of the $\dot{H}$ term they say that the underlying origin of the holographic dark energy is still unknown and that this term is contained also in the expression for RDE. Indeed for $\alpha=2\beta$ the suggested dark energy density reproduces the RDE density in a spatially flat spacetime.
Also note that for $\beta=0$, the above energy density reduces to that obtained by setting the Hubble radius as IR cut-off. Thus the new holographic dark energy density proposed above appears to be more general in form. On the other hand, contrary to the IR cut-off given by the event horizon, this model depends on local quantities,
avoiding in this way the causality problem.
In the following sections we attempt to investigate the effect of interaction of the new holographic dark energy with dark matter in a homogeneous and isotropic background from a general perspective. The function representing the interaction rate will be chosen to be a linear function of the density ratio so that we can recover results for different popular choices of interaction and also the no-interaction limit. The special cases {\it viz.} $\beta=0$, $\beta=2\alpha$ and $\alpha=1,\beta=2/3$ are discussed and desirable ranges for parameters range to avoid the phantom menace are also evaluated.

\section{The Interacting Model}
The field equations in Robertson-Walker spacetime filled with dark matter in the form of pressureless dust and dark energy are given by

\bea
3H^2+3\frac{k}{a^{2}} = \rho_{m}+\rho_{d}=(1+r)\rho_{d} \label{feqa}\\
 \mbox{and}\hspace{1cm}2\dot{H}+3H^2+\frac{k}{a^{2}} = -p_{d}= -w_{d}\rho_{d}.\label{feqb}
\eea
where $k$ denotes the spatial curvature, $\rho_{m}$ and $\rho_{d}$ are the respective energy densities of matter and the dark energy, $r\equiv\frac{\rho_{m}}{\rho_{d}}$ is the density ratio and $w_{d}\equiv\frac{p_{d}}{\rho_{d}}$ is the dark energy equation of state. Following the Bianchi identities the energy densities need to satisfy a conservation equation. Assuming an interaction between matter and dark energy the conservation equations read
\bea
\dot{\rho}_{m}+3H\rho_{m}=Q \label{conservation1}\\
    \mbox{and}\hspace{1cm}\dot{\rho}_{d}+3H(1+w_{d})\rho_{d}=-Q \label{conservation2}
\eea
where $Q>0$ implies that energy flows from dark energy to dark matter and for
$Q<0$ energy is transferred in the opposite direction. A vanishing $Q$ implies that matter and dark energy remain separately conserved. All these possibilities have been explored in literature. In a recent work Pereira and Jesus\cite{pere} have analyzed the possibility of dark matter decaying in to dark energy from a thermodynamical perspective. In the context of their model they have found that recent cosmological data favors the decay of dark matter in to dark energy over the inverse process. On the other hand generic conditions for the decay of dark energy in to fermionic fields have been studied by Macorra\cite{maco}. More recently Abdalla et al.\cite{abda} have suggested  a mechanism for the decay of dark energy in to dark matter where the dark energy is modeled by a scalar field with a special choice of potential.
\\
\par Introducing the dimensionless density parameters defined as
$$ \Omega_{m}\equiv\frac{\rho_{m}}{3H^{2}},~~~~~ \Omega_{d}\equiv\frac{\rho_{d}}{3H^{2}},~~~~~ \Omega_{k}\equiv-\frac{k}{a^{2}H^{2}}$$ and the deceleration parameter $q\equiv-(1+\frac{\dot{H}}{H^{2}})$, the field equations can be respectively recast as
\bea
\Omega_{d}(1+r)+\Omega_{k}=1\label{feq1}\\
\mbox{and}\hspace{1cm}q=\frac{1}{2}(1-\Omega_{k})\left(1+\frac{3w_{d}}{1+r}\right)\label{feq2}
\eea
The choice for the form of the interaction term $Q$ is however purely phenomenological owing to the lack of sound theoretical basis of its origin. Several forms have been put forward in literature\cite{zimdahl,fu,Qform,seta2} with $Q\propto H\rho$ being the most popular. Here $\rho$ represents the energy density of the dark sectors. A natural choice for the form of $Q$ is $\Gamma\rho_{d}$\cite{zimdahl} where $\Gamma$ determines the interaction rate which, in general varies with time.
Using the relation $\rho_d=3H^2\Omega_d$ and the above choice for $Q$ in the conservation equation (\ref{conservation2}) leads to
\be \label{omegadot}
\frac{\dot{\Omega}_{d}}{H\Omega_{d}}+2\frac{\dot{H}}{H^{2}}+3(1+w_{d})+\frac{\Gamma}{H}=0
\ee
which on further simplification using field equations (\ref{feqa}) and (\ref{feqb}) gives
\be \label{wd}
w_{d}=-\frac{1}{3(1-\Omega_{d})}\left[\frac{\dot{\Omega}_{d}}{H\Omega_{d}}+\Omega_{k}+\frac{\Gamma}{H}\right]
\ee
In view of equation (\ref{feq2}) the deceleration parameter $q$, therefore reads
\be \label{dp}
q=\frac{1}{2}-\frac{\Omega_{d}}{2(1-\Omega_{d})}\left[\frac{\dot{\Omega}_{d}}{H\Omega_{d}}+\frac{\Omega_{k}}{\Omega_{d}}+\frac{\Gamma}{H}\right]
\ee
\par A parameter of interest in a dark energy model is the ratio of energy densities $r\equiv\frac{\rho_{m}}{\rho_{d}}$. The evolution of this ratio plays an important role in addressing the `coincidence problem'. The observed coincidence of the dark energy density and matter density of the universe in recent epochs demand that the ratio $r$ either become constant or vary extremely slowly as compared to the scale factor $`a'$ over a considerably long period in the course of evolution of the Universe. It follows from equation (\ref{feq1}), that the dynamics of the density ratio is determined by
\be \label{rdot2}
\dot{r}=-(1+r)\left[\frac{\dot{\Omega}_{d}}{\Omega_{d}}+ \frac{\dot{\Omega}_{k}}{1-\Omega_{k}}\right]
\ee
which can also be expressed as
\be \label{rdot3}
\dot{r}=-\left[2qH\frac{\Omega_{k}}{\Omega_{d}}+(1-\Omega_{k}) \frac{\dot{\Omega}_{d}}{\Omega_{d}^{2}}\right]
\ee

\par It is clearly evident from equation (\ref{rdot2}) that for dark energies with constant density parameter (i.e $\dot\Omega_{d}=0$), the evolution of the density ratio $r$ is determined only by the spatial curvature. In case the spatial curvature also vanishes $(\Omega_{k}=\dot{\Omega}_{k}=0)$, then the ratio remains constant throughout the evolution of the Universe.
On the other hand for dark energies having a time varying density parameter (i.e $\dot\Omega_{d}\neq0$), as in case of holographic RDE\cite{gao}, the ratio $r$ will not remain constant throughout the evolution even in flat space. For such dark energies the signature flip in $q$ may be achieved even without any interaction between matter and dark energy. The variation of the dark energy density (and hence the density ratio) itself can facilitate the transition. In fact it can be easily shown from equation (\ref{dp}) that irrespective of the spatial curvature
\be \label{gammar}
r-\frac{\dot{\Omega}_{d}}{H\Omega_{d}}-\frac{\Gamma}{H}=2\frac{1-\Omega_{d}}{\Omega_{d}}q.
\ee
Therefore the transition from deceleration to acceleration will require matter and dark energy to evolve in a manner such that ($\frac{\dot{\Omega}_{d}}{H\Omega_{d}}+\frac{\Gamma}{H}$) is less than $r$ in the past and becomes greater than $r$ only in recent epochs. An appropriate evolution of the density ratio may also address the coincidence problem in addition.
\par The above set of equations(\ref{feqa}-\ref{rdot3}) give the most general description of the dynamics of the Universe in a homogenous isotropic background regardless of its spatial curvature, form of dark energy or its nature of interaction with matter.
In the following section we analyze in further detail the dynamics of the Universe for a specific choice of dark energy density assuming an interaction between dark matter and dark energy varying in a manner
\be
\frac{\Gamma}{H}=mr+n \label{gammah}
\ee
where $m$ and $n$ are constant parameters. Note that this simple choice of $\frac{\Gamma}{H}$ reproduces the various popular forms of $Q$\cite{fu} for different values of the parameters $m$ and $n$. Say for $m=0$ and $n\neq0$, $Q\propto H\rho_d$ whereas for $n=0$ but $m\neq0$, $Q\propto H\rho_m$. Again for $m=n\neq0$, $Q\propto H(\rho_m+\rho_d)$. Setting both $m$ and $n$ equal to zero, gives the no interaction limit.

\section{The new holographic dark energy interacting with matter}
\par Following the proposition by Granda and Oliveros\cite{granda} of a new infrared cut-off motivated purely out of dimensional considerations we introduce the holographic dark energy density of the form,
\be
\rho_{d}= 3(\alpha H^2 + \beta \dot {H})\label{rhox}
\ee
where $\alpha$ and $\beta$ are arbitrary non zero positive constants and $c=8\pi G=1$ is assumed.
By definition, it straight away follows that
\be \label{alpha}
\alpha=\Omega_{d}+\beta(1+q)
\ee
\par Note that $\beta=0$ implies $\rho_{d}\propto H^{2}$ and hence the dark energy density parameter $(\Omega_{d}=\alpha)$ shows no dynamical behavior. Such variation of the energy density is a characteristic of holographic dark energy with the infrared cutoff length set by the Hubble radius. As has been discussed in the previous section, for such dark energies the evolution of the density ratio $r$ is determined solely by the spatial curvature and in flat space the ratio remains constant throughout the evolution of the Universe [{\it see equation} (\ref{rdot2})]. It is also interesting to note from (\ref{dp}) that a spatially flat Universe filled with the above form of dark energy will never undergo a transition from decelerated to accelerated phase of expansion unless the dark energy interacts with the matter component. In fact in the absence of any such interaction in flat space, the assumed dark energy component will only mimic the pressureless matter [{\it see equation} (\ref{wd})] thereby failing to generate any negative pressure to drive the cosmic acceleration. From equation (\ref{gammar}) which is true for all three cases $k=0,\pm1$, it is clearly evident that since $\dot{\Omega}_d=0$, the transition dynamics of the Universe under the influence of holographic dark energy with Hubble scale cutoff is completely determined by the difference of the two ratios $r$ and $\frac{\Gamma}{H}$. The desired signature change in $q$ requires the difference ($r-\frac{\Gamma}{H}$) to evolve from positive to negative in the course of evolution of the Universe. Thus in a spatially flat model where $r$ remains constant, the signature flip in $q$ will be realized purely as an interaction effect. A detailed analysis of interacting holographic dark energy with cutoff scale $H^{-1}$ is available in \cite{zimdahl}.

\par It has been already argued that a general description of any dark energy model must
allow an interaction between the two dark sectors. For a non zero $\beta$, assuming an interaction between dark matter and the new holographic dark energy as prescribed in (\ref{gammah})
and making use of equations (\ref{dp}) and (\ref{feq1}) in equation (\ref{alpha}) one gets,
\be \label{deq1}
\frac{\dot{\Omega}_{d}}{H\Omega_{d}}=\frac{(1-\Omega_{d})(2\Omega_{d}-2\alpha+3\beta)-(n-m)\beta\Omega_{d}-(1-m)\beta\Omega_{k}-m\beta}{\beta\Omega_{d}}
\ee

One may in principle attempt to solve the above differential equation for different values of the spatial curvature $k$. However analytical solution of the equation become simple for vanishing spatial curvature. In the following analysis, we assume the Universe to be spatially flat which implies $\Omega_{k}=0$ and therefore equation (\ref{deq1}) reduces to
\be \label{deq2}
\beta\dot{\Omega}_{d}=-(2\Omega_{d}^{2}+B\Omega_{d}+C)H
\ee
where $B$ and $C$ are constants defined as
$$B\equiv(n-m)\beta+3\beta-2\alpha-2$$ and $$C\equiv m\beta-3\beta+2\alpha .$$ Expressing the Hubble parameter, in terms of the redshift $z$ as $ H=-\frac{\dot{z}}{1+z}$ the equation (\ref{deq2}) can be easily integrated for $B^{2}>8C$ to obtain
\be \label{Omegad}
\Omega_{d}(z)=\frac{A(B+\xi)(1+z)^{\frac{\xi}{\beta}}-B+\xi}{4\left[1-A(1+z)^{\frac{\xi}{\beta}}\right]}
\ee
Here, $A$ is a constant of integration and $\xi$ is simply a shorthand notation used to represent $\sqrt{B^{2}-8C}$ which here is a real positive quantity.

\par Determination of the deceleration parameter $q$ and the dark energy equation of state $w_{d}$ is now straight forward. Substituting (\ref{Omegad}) for $\Omega_{d}$ in (\ref{alpha}), the deceleration parameter $q$ in terms of the redshift $z$ reads
\be \label{qz}
q(z)=\frac{A(B+\xi)(1+z)^{\frac{\xi}{\beta}}-B+\xi}{4\beta[A(1+z)^{\frac{\xi}{\beta}}-1]}+\frac{\alpha}{\beta}-1
\ee
On the other hand in view of the field equations (\ref{feq1}) and (\ref{feq2}), for a spatially flat Universe,
\be \label{q}
2q=1+3w_{d}\Omega_{d}
\ee
and therefore using (\ref{Omegad}) and (\ref{qz}) in the above we easily get
\be \label{wz}
w_{d}(z)= \frac{2}{3\beta}\left[\frac{(4\alpha-6\beta)[1-A(1+z)^{\frac{\xi}{\beta}}]}{A(B+\xi)(1+z)^{\frac{\xi}{\beta}}-B+\xi}-1\right]
\ee
The ratio of densities $r$ is obtained for $\Omega_{k}=0$, using equation (\ref{feq1}),
\be \label{rz}
r(z)=\frac{1}{\Omega_{d}}-1= \frac{4\left[1-A(1+z)^{\frac{\xi}{\beta}}\right]}{A(B+\xi)(1+z)^{\frac{\xi}{\beta}}-B+\xi}-1
\ee

\par From the above expressions it is clear that the evolution of the Universe in this model is determined by the choice of the constant parameters $\alpha$, $\beta$, $m$ and $n$. Imposing appropriate observational constraints, one can fix these parameters to obtain a viable dark energy model. Current observational data suggest that at the present epoch, the dark energy density parameter $\Omega_{d0}= 0.73$, whereas the dark energy equation of state $w_{d0}$ has a value close to $-1$. Observations also suggest that the transition of the Universe to an accelerated phase of expansion from a decelerated phase has occurred very recently at a redshift $0.3<z_{t}<1.2$. At this point one may note that the value of the deceleration parameter at the present epoch is automatically determined by the present values of $\Omega_{d}$ and $w_{d}$ due to equation (\ref{q}) and therefore assuming $\Omega_{d0}=0.73$ and $w_{d0}=-1$, the present value of the deceleration parameter will be fixed at $q_{0}=-0.595$ irrespective of the choice of dark energy and its rate of interaction with matter.
\par Now it follows from equation (\ref{Omegad}) that the value of $\Omega_{d}$ at the present epoch $(z=0)$ is
\be\label{Omegad0}
\Omega_{d0}=\frac{\xi(1+A)-B(1-A)}{4(1-A)}
\ee
which implies that the constant of integration will be given by
\be\label{A}
A=\frac{4\Omega_{d0}+B-\xi}{4\Omega_{d0}+B+\xi}
\ee
Setting $w_{d}=w_{d0}$ at $z=0$ in equation (\ref{wz}) and simplifying using relation (\ref{A}) determines the parameter $\alpha$ in terms of $\beta$ given by
\be \label{alpha1}
\alpha=\frac{3}{2}\beta(1+w_{d0}\Omega_{d0}) + \Omega_{d0}
\ee
Interestingly the presence or absence of interaction does not have any bearing on the above relation between $\alpha$ and $\beta$.
Finally, setting $q=0$ at $z=z_{t}$, equation (\ref{qz}) leads to
\be \label{zt}
(1+z_{t})^{\frac{\xi}{\beta}}=\frac{4\beta-4\alpha-B+\xi}{A(4\beta-4\alpha-B-\xi)}
\ee
Replacing $A$, $\xi$, $B$ and $C$ in terms of $\alpha$, $\beta$, $m$ and $n$ yields a relation between these constant parameters of the model with various physical quantities of interest. We discuss below the effect of the choice of these parameters on the dynamics of the Universe and other relevant physical quantities.
\subsection{\small Holographic Ricci dark energy model}
\par Let us first focus on the special case $\alpha=2\beta$, which generates the holographic Ricci dark energy with density proportional to the Ricci scalar $R$\cite{gao}. Equation (\ref{rhox}) gives
\be \label{rde}
\rho_d=3\beta(2H^2+\dot{H})
\ee
Eliminating $\alpha$ from (\ref{alpha1}) and solving for $\beta$ with $\Omega_{d0}=0.73$ and $w_{d0}=-1$ we get $\beta\simeq 0.458$ and therefore we have $\alpha\simeq0.915$. For no interaction between matter and the dark energy the transition from decelerated to accelerated expansion occurs at a redshift $z_t=0.55$ as obtained from (\ref{zt}). The evolution of the dark energy equation of state $w_d$ depicted in Fig.1a clearly show that $w_d$ is nearly zero in the past and approaches $-1.12$ in the distant future. Thus in this model the Universe crosses in to the phantom era in future thereby showing a `quintom' like behavior. Similar results have been obtained in \cite{gao} as well where non interacting RDE model was studied. It may be pointed out here that if one does not fix by hand the present value of the dark energy equation of state $w_{d0}$ then the parameter $\beta$ will remain free. A plot of $z_t$ against $\beta$ (Fig.1b) show that for the desired transition to occur at $z_t>0.3$, $\beta$ must lie in the range $0.145\leq\beta\leq0.634$.
It is indeed interesting to note that an entry in to the phantom regime can be avoided in this model for $\beta\geq0.5$ [{\it see} Fig.1c]. Since the violation of the positive energy condition is incompatible with the holographic idea therefore the allowed range of $\beta$ may further be restricted to $0.5\leq\beta\leq0.634$. With a value of $\beta$ with in this range $q_0$ and $w_{d0}$ may have values in the range $-0.15\geq q_0\geq-0.46$ and $-0.595\geq w_{d0}\geq-0.877$ respectively. Thus higher the value of $\beta$ in the allowed range, the closer the present value of dark energy equation of state to -1. Fig.1d shows the variation of the ratio of energy densities $r$ against $z$ for $\beta=0.5$. The ratio $r$ remains almost constant at a value slightly greater than $2$ over a long period of evolution of the Universe and decreases to drop below $1$ only in the recent past. This reflects that matter and dark energy densities has almost been of the same order through a sufficiently long period of evolution of the Universe. The coincidence problem therefore does not arise in this model. It is to be seen whether the presence of interaction can produce an even more slowly varying function $r(z)$ and further improve the results.

\begin{figure}[h]
\begin{center}
\includegraphics[width=6cm]{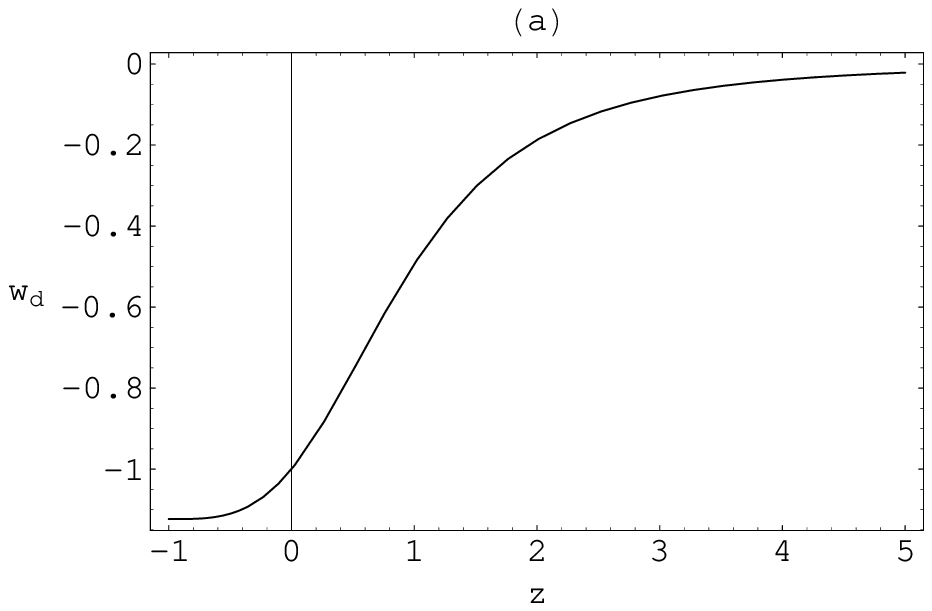}
\includegraphics[width=6cm]{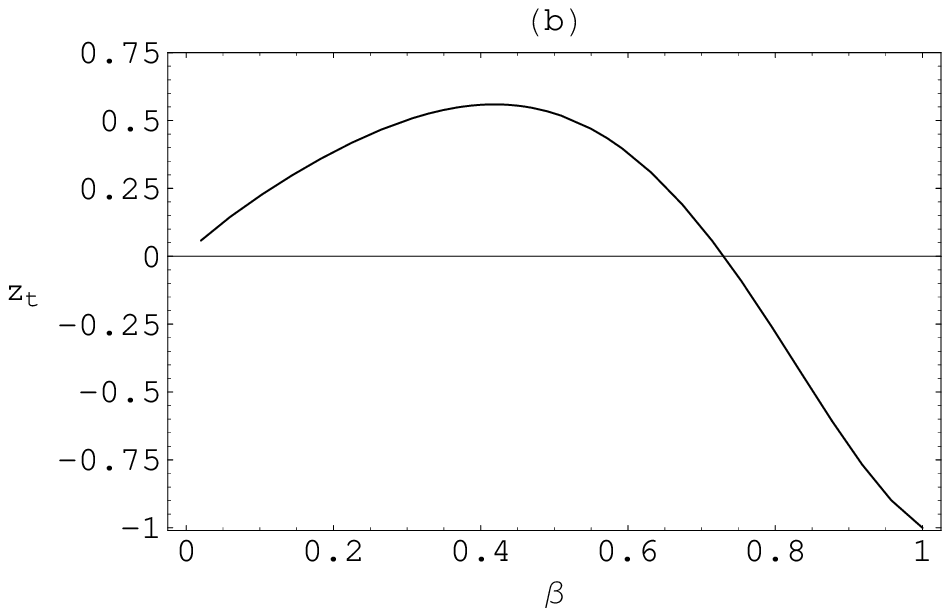}
\includegraphics[width=6cm]{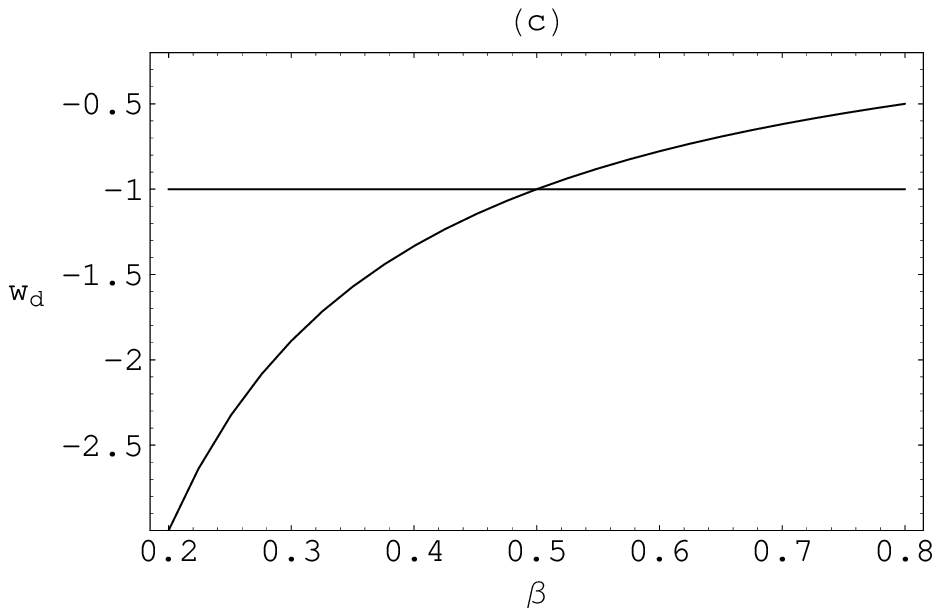}
\includegraphics[width=6cm]{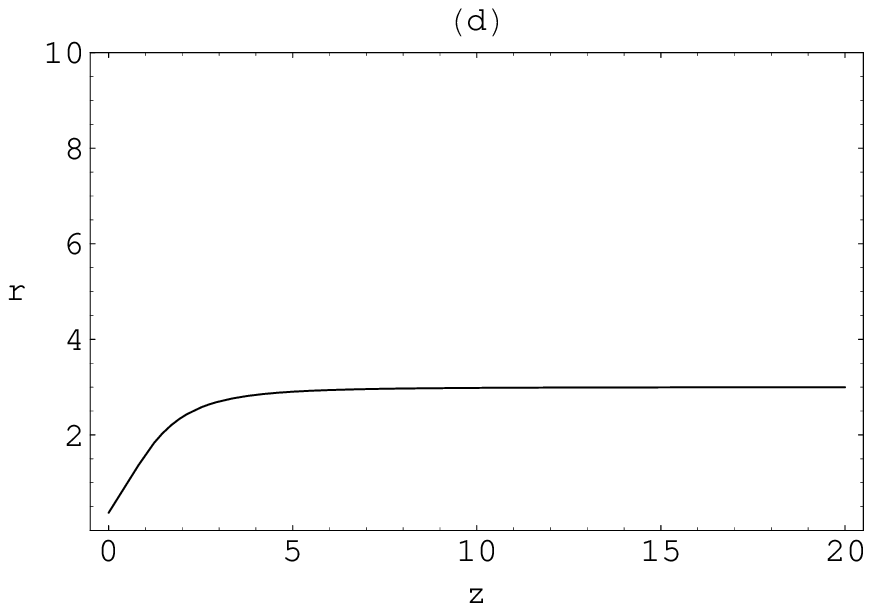}
\caption{(a)\it Plot of $w_d$ vs $z$ for $\beta=0.458$ and $w_{d0}=-1$; (b) Plot shows how the transition redshift $z_t$ changes with the choice of $\beta$; (c)$w_d$ vs $\beta$ at $z=-1$ showing that for $\beta\geq0.5$ $w_d\geq-1$ and (d) $r$ vs $z$ for $\beta=0.5$. In all the plots $\alpha=2\beta$, $m=n=0$ and $\Omega_{d0}=0.73$.}
\end{center}
\end{figure}

\par Interacting holographic RDE model have been analyzed in detail in \cite{fu} where all three phenomenological forms of interaction namely $Q$ proportional to $H\rho_d$[IRDE1], $H\rho_m$[IRDE2] and $H(\rho_d+\rho_m)$[IRDE3] have been considered. The parameters of the models are evaluated using latest observational constraints from SN-Ia, CMBR, WMAP and BAO data. Note that a choice ($\beta=0.442$, $m=0$, $n=0.156$) in the present work corresponds to IRDE1 model of \cite{fu}. Similarly the correspondence with IRDE2 and IRDE3 models are respectively achieved by setting ($\beta=0.433$, $m=0.096$, $n=0$) and ($\beta=0.437$, $m=n=0.060)$). Needless to mention that $\alpha=2\beta$ in each case. Use of these values in addition to $\Omega_{d0}=0.73$ ({\it i.e} $\Omega_{m0}=0.27$) in equation (\ref{alpha1}) shows that in all three cases $w_{d0}<-1$ implying that the Universe is already being dominated by a phantom like dark energy. Like in case of non interacting RDE model, crossing the phantom barrier in any one of these interacting RDE models can also be avoided for $\beta\geq0.5$. Compared to the non interacting model the transition to accelerated expansion in all three interacting cases is found to occur at higher redshifts ($z_t=$ 0.6 to 0.7). The upper bound on the value of $\beta$ which ensures $z_t>0.3$ is found to be slightly higher than $0.634$ in each of the models. This apart there is nothing much that distinguishes the interacting RDE models among themselves or from the non-interacting one. The $r$ vs $z$ curves for the three interacting models almost coincides with the plot in fig. 1d. In other words the interacting RDE models does not seem to address the coincidence problem any better than the non interacting case.

\subsection{\small The variable $\Lambda$ equivalent model}
Another special choice of parameters in the new holographic dark energy density corresponds to $\alpha=1$ and $\beta=2/3$. Equation (\ref{wz}) shows that for the above choice the dark energy equation of state remains constant at $-1$ at all redshifts regardless of whether the dark energy interacts with matter or not. The model therefore automatically avoids the phantom menace and the dark energy density in this case effectively plays the role of a variable $\Lambda$ parameter. If we claim this dark energy to be of holographic origin, then its IR cut off length may be identified with $\Lambda^{-1/2}$. The cosmological parameter $\Lambda$, if it exists, may indeed be a natural choice for the IR cutoff.
By virtue of equation (\ref{q}) the value of deceleration parameter at the current epoch in this model will therefore be $q_0=-0.595$ for $\Omega_{d0}=0.73$. In absence of any interaction (i.e $m=n=0$) there is no free parameter in this model and all physical quantities are uniquely determined. The transition to accelerated phase occurs at redshift $z_t=0.755$ and from (\ref{qz}) we find that $q(z)=1/2$ at large $z$, a condition ambient for large scale structure formation. If the dark energy is allowed to interact with matter in a manner such that $m=0$ but $n\neq0$ then from equation (\ref{zt}) it is found that for the transition to occur within a desired redshift range the interaction parameter $n$ must take a value within $-1.55<n<0.362$. On the other hand for interaction of the form $n=0$ $m\neq0$, $m$ must be positive definite since $m<0$ leads to $\Omega_d(z)<0$ at large $z$ which is physically unviable. The value of $m$ that would produce a signature flip in $q$ at an appropriate redshift lie in the range $0<m<0.428$. For interaction of the third kind where $m=n\neq0$, these parameters need to be positive with an upper bound of 0.201 for giving feasible results. The deceleration parameter $q$ remains constant at 1/2 in the far past for the first form of interaction whereas in the latter two cases $q$ at large $z$ is constant at a value little lower than 0.5. Higher the value of $m$ lower is the value of $q$ at large $z$. Fig. 2 below shows the variation of $r$ against $z$ for different forms of interaction. The plots point out that the model under consideration is plagued by the `coincidence problem'. It is however observed from the plots that for $n=0$ but $m\neq0$ the ratio $r$ remains almost constant at a sufficiently low value over a very long period of evolution of the Universe and varies slowly only in the recent past. It may therefore be claimed that an interaction of the form $Q\propto H\rho_m$ is relatively more suitable in avoiding the coincidence problem compared to the non interacting model and the other two interacting models.

\begin{figure}[h]
\begin{center}
\includegraphics[width=6cm]{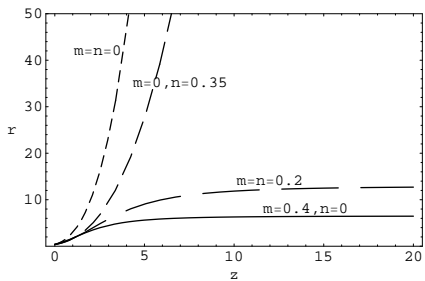}
\caption{$r$ vs $z$ plot for $\alpha=1$, $\beta=2/3$ and $\Omega_{d0}=0.73$ and different choices of $m$ and $n$.}
\end{center}
\end{figure}

\section{Conclusions}

In the present work we have investigated the evolution of a homogeneous and isotropic Universe with a special choice of the dark energy density $\rho_d=3(\alpha H^2+\beta \dot{H})$ inspired by the cosmological Holographic Principle. The dark energy is allowed to interact with matter at a rate which is chosen to be a linear inhomogeneous function of $r$, the ratio of energy densities of dark matter and dark energy. Such a form enables us to reproduce all the three popular phenomenological forms of interaction existing in literature, for different choices of the parameters. The equations arrived at are very general in nature based on which three special cases have been discussed.

\par For $\beta=0$, the dark energy density becomes proportional to $H^2$ reducing thereby the infrared cut-off scale to Hubble length. The evolution of the density ratio $r$ in this model is determined only by the spatial curvature. For a vanishing curvature this ratio remains constant throughout the evolution of the universe solving the `coincidence problem', but the model never achieves a transition from decelerated to accelerated phase of expansion unless dark energy interacts with matter.

\par The choice ($\alpha=2\beta$) gives the holographic Ricci dark energy model in which the dark energy density becomes proportional to the Ricci scalar $R$. Any dark energy of holographic origin requires that its equation of state $w_d$ must not be phantom like\cite{zimdahl,bak}. In previous works on non interacting and interacting holographic RDE models\cite{gao,fu}, it is found that the dark energy equation of state crosses the phantom divide for suggested values of the model parameters. In the present work it has been shown that for $\Omega_{d0}=0.73$ a choice of $\beta>0.5$ can restrict the RDE model from crossing the phantom barrier both in the absence and presence of interaction. The $r$ vs $z$ plot reveal that matter and dark energy densities were almost of the same order for over a long period of evolution. So the coincidence problem is thereby avoided. The presence of interaction in this model however does not seem to substantially alter the dynamics of the non interacting model. Therefore it is indeed difficult to make a choice of preference between non interacting and interacting RDE models.

\par Finally the choice $\alpha=1$,$\beta=2/3$ have been worked out. This choice results in a constant dark energy equation of state $w_d=-1$, and therefore is equivalent to a varying $\Lambda$ model. From holographic dark energy point of view this implies considering $\Lambda^{-1/2}$ as the IR cut off length. For $\Omega_{d0}=0.73$ the values of $z_t$ and $q_0$ in this case remain fixed at 0.755 and -0.595 respectively, regardless of the nature of interaction. However the presence of interaction in this model is found to play a significant role in addressing the coincidence problem. The allowed range of the interaction parameters for producing observationally viable results have been estimated for each different form of interaction. It has been concluded that a choice of interaction of the form $Q\propto H\rho_m$ may be better suited to address the cosmic coincidence problem in this special case of the new holographic dark energy model.\\
\\
{\bf Acknowledgements:}
\par The authors are grateful to Prof. N. Banerjee of IISER-Kolkata; India, for his valuable comments and suggestions.

\end{document}